\documentclass[preprintnumbers,secnumarabic,amsmath,amssymb,nofootinbib,floatfix]{revtex4}
\usepackage{varioref,exscale,latexsym,amsmath,amssymb}
\usepackage{graphicx}
\usepackage[colorlinks]{hyperref}

\usepackage{graphicx}
\usepackage{slashed}
\usepackage{dcolumn}
\usepackage{bm}
\newcommand{\beq}{\begin{equation}}
\newcommand{\eeq}{\end{equation}}
\newcommand{\bea}{\begin{eqnarray}}
\newcommand{\eea}{\end{eqnarray}}
\newcommand{\nn}{\nonumber}

\def\lsi{\raise0.3ex\hbox{$<$\kern-0.75em\raise-1.1ex\hbox{$\sim$}}}
\def\gsi{\raise0.3ex\hbox{$>$\kern-0.75em\raise-1.1ex\hbox{$\sim$}}}

\def\beq{\begin{equation}}

\def\eeq{\end{equation}}

\def\beqa{\begin{eqnarray}}

\def\eeqa{\end{eqnarray}}

\begin{document}
\preprint{ACFI-T17-22}

\title{{\bf Inducing the Einstein action in QCD-like theories}}

\medskip\

\author{John F. Donoghue}
\email{donoghue@physics.umass.edu}
\affiliation{~\\
Department of Physics,
University of Massachusetts\\
Amherst, MA  01003, USA\\
 }

\author{Gabriel Menezes}
\email{gsantosmenez@umass.edu}
\affiliation{~\\
Department of Physics,
University of Massachusetts\\
Amherst, MA  01003, USA\\
}

\affiliation{~ Departamento de F\'{i}sica, Universidade Federal Rural do Rio de Janeiro, 23897-000, Serop\'{e}dica, RJ, Brazil \\
 }

\begin{abstract}
We evaluate the induced value of Newton's constant which would arise in QCD. The ingredients are modern lattice results, perturbation theory and the operator product expansion. The resulting shift in the Planck mass is positive. A scaled-up version of such a theory may be part of a quantum field theory treatment of gravity.
\end{abstract}
\maketitle

\section{Introduction}

The action for a pure Yang-Mills theory, such as the gluonic sector of QCD, is scale invariant. Nevertheless, a scale enters the theory through the running coupling constant, which defines an energy scale at which the coupling becomes large. The spectrum and observables of the theory will depend on this scale through dimensional transmutation.

In particular, when we include the metric as a field as we do in general relativity, QCD will induce a change in the cosmological constant and in the gravitational constant $G$. Since the cosmological constant is related to the energy density of the vacuum, the QCD contribution to it has a simple expression in terms of the trace of the energy-momentum tensor
\begin{equation}\label{cosmconst}
4 \Lambda_{\textrm{ind}} =  \langle 0| T^\mu_{~\mu}|0\rangle  \  \
\end{equation}
where $|0\rangle$ is the vacuum. The shift in the gravitational constant is given by the Adler-Zee formula \cite{Adler:1980pg, Zee:1980sj, Zee:82, Adler:1982ri, Sakharov:1967pk} (to be reviewed in Sec. 2) in terms of the correlation function
\begin{equation}
\frac{1}{16\pi G_{\textrm{ind}}} = \frac{i}{96}\int d^4x ~x^2~
\langle 0|T( T^\mu_{~\mu}(x)T^\nu_{~\nu}(0)|0\rangle  \ \ .
\end{equation}
In this paper we provide a determination of the induced gravitational constant defined by the Adler-Zee formula in QCD.

In gluonic QCD, the trace of the energy momentum tensor is given by the trace anomaly
\begin{equation}
T^\mu_{~\mu} = \frac{\beta(g)}{2g} F^a_{\mu\nu}F^{a\mu\nu}  \ \
\end{equation}
where $\beta(g)$ is the renormalization-group beta function of QCD and $g$ is the (dimensionless) coupling constant. The Adler-Zee formula involves the correlation function of $F^2$, which has been studied in the context of scalar glueballs. The ingredients needed for the evaluation are then perturbation theory \cite{Adler:1982ri} and the OPE expansion \cite{Novikov:1979va, Novikov:1983gd, Bagan:1990sy} at short distance and modern lattice glueball studies \cite{Chen:2005mg}  at long distances. We match these contributions at an distance/energy scale $x^{-1}= X_0^{-1}=2$ GeV (in units of $\hbar = c = k_{B} = 1$, which will be consistently employed throughout the paper). While there is some residual matching dependence, this procedure determines that the induced $G$ is positive and evaluates its magnitude to within about 30\%.

While this calculation can be considered as simply a demonstration of a feature of QCD, there is potentially another motivation in gravitational physics. Strongly interacting theories similar to QCD could occur at higher energies also. There would be shifts in the gravitational constant also in such theories also. Perhaps the effective Planck mass
\begin{equation}
M_P^2 = \frac{1}{G}
\end{equation}
is in fact determined by the strongly interacting theory with the largest intrinsic scale. This would occur if the coefficient of the Einstein action in the ultimate theory of quantum gravity was smaller than the Yang-Mills scale or absent altogether, as would happen in scale/confomally invariant gravitational theories  \cite{Stelle:1976gc, Fradkin:1981hx, Donoghue:2016vck, Holdom:2015kbf}. So perhaps the Planck scale is a manifestation of a high scale Yang-Mills theory. The fact that the QCD result comes out to be positive is important for such a possibility. We do not analyse such gravitational theories in this paper, but we plan to return to that topic in future work.

The outline of the paper is as follows. In Section 2 we briefly review the origin of the Adler-Zee formula. In the subsequent section, we collect the various ingredients for the evaluation. Section 4 is devoted a numerical evaluation as well as a discussion of the uncertainties. In Sec. 5 we provide a summary. The Appendix is devoted to reconciling a (previously unnoticed) discrepancy in the literature involving a related sum-rule formula for the cosmological constant, where the works of Novikov et al \cite{Novikov:1979va} and of Brown and Zee \cite{Brown:1982am} yield sum-rules that differ by a factor of two.

\section{Brief review of the Adler-Zee formula}

Here we review the induced gravitational effects due to the matter sector of a quantum field theory coupled to the metric. We take the Minkowski metric as $\eta_{\mu\nu} = \textrm{diag}(1,-1,-1,-1)$ and the Riemann curvature tensor given by
$R^{\lambda}_{\ \mu\nu\kappa} = \partial_{\kappa}\Gamma^{\lambda}_{\mu\nu} + \Gamma^{\eta}_{\mu\nu}\Gamma^{\lambda}_{\kappa\eta} - (\nu \leftrightarrow \kappa)$. We define the gravitational effective action by
\beq
e^{iS_{\textrm{eff}}[g_{\mu\nu}]} = \int d\phi e^{i S_{\textrm{matter}}[\phi,g_{\mu\nu}]},
\eeq
where $\phi$ represents generically the matter fields and $S_{\textrm{matter}}[\phi,g_{\mu\nu}]$ describes matter fields on a curved background. The action $S_{\textrm{eff}}[g_{\mu\nu}]$ is a scalar under general-coordinate transformations. This observation allows one to represent it as the integral over the manifold of a scalar density. Formally, for slowly varying metrics, one has the following series expansion in powers of $\partial_{\lambda}g_{\mu\nu}$
\bea
S_{\textrm{eff}}[g_{\mu\nu}] &=& \int d^{4}x \sqrt{-g} {\cal L}_{\textrm{eff}}[g_{\mu\nu}]
\nn\\
{\cal L}_{\textrm{eff}}[g_{\mu\nu}] &=& {\cal L}^{(0)}_{\textrm{eff}}[g_{\mu\nu}] + {\cal L}^{(2)}_{\textrm{eff}}[g_{\mu\nu}]
+ {\cal O}[(\partial_{\lambda}g_{\mu\nu})^{4}]
\nn\\
{\cal L}^{(0)}_{\textrm{eff}}[g_{\mu\nu}] &=& - \Lambda_{\textrm{ind}},\,\,\,
{\cal L}^{(2)}_{\textrm{eff}}[g_{\mu\nu}] = \frac{R}{16\pi G_{\textrm{ind}}}.
\eea
Our task is to derive representations for the induced cosmological constant $\Lambda_{\textrm{ind}}$ and the induced Newton's gravitational constant $G_{\textrm{ind}}$ in terms of the vacuum expectation value of products of the stress-energy tensor $T_{\mu\nu}$ of the matter fields. For a discussion on the conditions that $\Lambda_{\textrm{ind}}$ and $G_{\textrm{ind}}$ should satisfy in order to be uniquely calculable in terms of the renormalized parameters of the flat space-time matter theory, see the review~\cite{Adler:1982ri}. Here we consider a matter Lagrangian - in our case QCD - coupled covariantly to the gravitational field. For weak fields the metric can be expanded using $g_{\mu\nu} = \eta_{\mu\nu} + h_{\mu\nu}$. One finds

\begin{equation}\label{expansion}
{\cal L}_{\textrm{matter}}[g_{\mu\nu}, A_\mu] = {\cal L}_{0}[\eta_{\mu\nu}, A_\mu] - \frac12 h^{\mu\nu} T_{\mu\nu}
+ \frac14 h^{\mu\nu}h^{\alpha\beta}\tau_{\mu\nu,\alpha\beta} +...  \ \ .
\end{equation}
On the right hand side of this equation all the indices are raised and lowered with the flat space metric. The term linear in $h^{\mu\nu}$ yields the energy-momentum tensor and there is a term quadratic in $h^{\mu\nu}$ also. The latter will be discussed in the Appendix as it plays a role in the cosmological constant sum rule.

The effective action for the gravitational field will then include contributions from the interactions of the matter fields. Expansion in powers of the field $h^{\mu\nu}$ yields
\begin{eqnarray}
iS_{\textrm{eff}}[h] &=& -\frac{i}{2} \int d^{4}x h^{\mu\nu}(x) \langle T_{\mu\nu}(x) \rangle
+ \frac{i}{4} \int d^{4}x h^{\mu\nu}(x)h^{\alpha\beta}(x) \langle \tau_{\mu\nu,\alpha\beta}(x) \rangle
\nn\\
&+& \frac{i^{2}}{2!}\left(\frac{1}{2}\right)^{2} \int d^{4}x \int d^{4}y\,
h^{\mu\nu}(x)h^{\rho\sigma}(y)\langle T \{{\bar T}_{\mu\nu}(x){\bar T}_{\rho\sigma}(y)\}\rangle
+ {\cal O}(h^3)
\end{eqnarray}
where $\langle \ldots \rangle = \langle 0| \ldots |0 \rangle$ denotes vacuum expectation value and ${\bar T}^{\mu\nu}(x) = T^{\mu\nu}(x) - \langle T^{\mu\nu}(x) \rangle$. The above expression is very similar to the usual expansion for the generating functional of connected correlation functions in quantum field theory, as long as one envisages $h_{\mu\nu}(x)$ as an external
field. As alluded above, the extra contribution coming from the second term on the right-hand side is necessary for consistency.

Following Zee \cite{Zee:1980sj}, at this stage it is most convenient (but not required \cite{Brown:1982am}) to specialize to the trace of the metric, using $h_{\mu\nu}(x) = \frac14 \eta_{\mu\nu} h(x)$. In this case the action only involves the trace of the energy momentum tensor, $T(x) = \eta_{\mu\nu} T^{\mu\nu}(x) $. For long wavelength metric fields, in our case wavelengths longer than the QCD scale, we can Taylor expand the metric
\begin{equation}
h(y) = h(x)+ (y-x)^\mu \partial_\mu h(x) +\frac12 (y-x)^\mu (y-x)^\nu \partial_\mu \partial_\nu h (x) +.... \ \ .
\end{equation}
The effective Lagrangian can then be identified as
\begin{eqnarray}
i{\cal L}_{\textrm{eff}}[h] &=& -\frac{i}{2}  h(x) \left[\frac{1}{4}\langle T(x) \rangle\right]
+ \frac{i}{16}(h(x))^{2}\left[ \frac{1}{4} \eta^{\mu\nu}\eta^{\rho\sigma}\langle \tau_{\mu\nu,\rho\sigma}(x) \rangle \right]
\nn\\
&-&\, \frac{1}{16} (h(x))^{2}\int d^{4} z \left[\frac{1}{8} \langle T \{{\bar T}(z){\bar T}(0)\}\rangle\right]
+ \frac{1}{2^{10}}(\partial_{\mu} h)^{2}\int d^{4}z\,z^2\left[\langle T \{{\bar T}(z){\bar T}(0)\}\rangle\right]
+ {\cal O}(h^3).
\label{Leff}
\end{eqnarray}
The terms without any derivatives of the metric are related to the cosmological constant. In particular the term linear in $h$ reproduces Eq.~(\ref{cosmconst}). The Einstein action involves two derivatives of the metric. For this form of the metric trace we have
\begin{equation}
\sqrt{-g} R = - \frac{3}{32} (\partial_\mu h(x))^2 +... \  \ .
\end{equation}
Comparing the induced Lagrangian with the Einstein-Hilbert Lagrangian, one finds the Adler-Zee formula describing the QCD contribution to the Einstein-Hilbert action
\begin{equation}
\frac{1}{16\pi G_{\textrm{ind}}} =  \frac{i}{96} \int d^{4}z\,z^2 \langle T \{{\bar T}(z){\bar T}(0)\}\rangle.
\end{equation}
Notice that this sum-rule involves the vacuum-subtracted version of the energy-momentum tensor, and we will treat this as being implied in subsequent work.

The Adler-Zee formula involves the two-point function of the trace of the energy momentum tensor, which we will follow Adler and call $\psi(x)$
\begin{equation}
\psi(x) =  \langle T \{{\bar T}(x){\bar T}(0)\}\rangle
\end{equation}
The Fourier transform of this gives the momentum-space correlator
\begin{equation}
\Pi_{AZ} (q^2) = -i \int d^4 x e^{iq\cdot x}   \langle T \{{\bar T}(x){\bar T}(0)\}\rangle
\end{equation}
In terms of the correlator, the induced gravitational constant involves the first derivative at zero momentum
\begin{equation}\label{momentumG}
\frac{1}{16\pi G_{\textrm{ind}}} = \frac{1}{12} \Pi_{AZ}'(0)   \ \
\end{equation}
In the course of evaluating the induced value of $G$, we will work in Euclidean space. The corresponding formulas there are
\begin{equation}
\frac{1}{16\pi G_{\textrm{ind}}} = -\frac{1}{96} \int d^4y_E~ y^2   \langle T \{{\bar T}(y){\bar T}(0)\}\rangle \ \
\end{equation}
and
\begin{equation}\label{momentumGEuc}
\frac{1}{16\pi G_{\textrm{ind}}} = -\frac{1}{12} \frac{d}{dQ^2}\Pi_{AZ}(Q^2)|_{Q^2=0}  \ \
\end{equation}

\section{Ingredients to the sum-rule}

The integration in the Adler-Zee formula runs over all distances. At long distances we are unable to calculate analytically. However, this particular correlation function is related to one which has been used to determine glueball properties, and has been studied on the lattice. We will use the most recent lattice work which yields the parameters which we will need \cite{Chen:2005mg}. However the lattice studies do not probe the shorter distance properties. At the shortest distance, the perturbative contributions have been calculated by Adler \cite{Adler:1982ri}. In the intermediate energy range, there are QCD sum-rule techniques, dating back to the work of Novikov et al (NSVZ) \cite{Novikov:1979va}, which use the operator product expansion (OPE) to describe some contributions which are subleading to the perturbative contribution but still relevant at moderate energies. We will seperate the problem into the long and short distance contributions. We tie them together at a distance which correspond to an energy of 2 GeV.

After performing a change of variables $x^{2} = t$, we split the integration into an ultraviolet part and an infrared part as follows:
\bea
\frac{1}{16\pi G_{\textrm{ind}}} &=& - \frac{\pi^2}{96}(I_{UV} + I_{IR})
\nn\\
I_{\textrm{UV}} &=& \int_{0}^{t_{0}} dt t^{2} \Psi(t)
\nn\\
I_{\textrm{IR}} &=& \int^{\infty}_{t_{0}} dt t^{2} \Psi(t).
\label{UV-IR}
\eea
The high-energy portion $I_{\textrm{UV}}$ contains perturbative contributions coming from short-distance scales as well as terms coming from intermediate energies which will be assessed through a operator product expansion technique as mentioned above.

First let us discuss the infrared part $I_{\textrm{IR}}$. As discussed previously, it will be estimated within lattice methods. For large Euclidean $x$, one considers that $\Psi$ takes the form of the correlation function for a massive scalar particle:
\beq
\Psi_{\textrm{IR}}(x) = \frac{\lambda^{2}}{4\pi^{2}} \frac{M_g}{x} K_{1}(M_g x)
= \lambda^{2} \int \frac{d^{4}p}{(2\pi)^4}\,\frac{e^{i {\bf p} \cdot {\bf x} + i p_{4}\tau}}
{{\bf p}^2 + p_{4}^{2} + M_g^{2}},
\label{infrared}
\eeq
where $K_{1}(z)$ is a modified Bessel function, $x = \sqrt{|{\bf x}|^2 + \tau^2}$ ($\tau$ is the ``Euclidean time"), $M_g$ is the glueball mass and
\beq
\lambda = \langle 0| {\bar T}(0)| S\rangle   \ \ ,
\eeq
is the glueball coupling, with $| S\rangle$ being the normalized scalar glueball state.

For the high-energy/short-distance component, $I_{\textrm{UV}}$, we start with the perturbative contribution.  It can be calculated directly in position space and where the form is~\cite{Adler:1982ri}
\beq
\psi_{\textrm{pert}} =\frac{ C_\psi}{x^8 (\log (1/\Lambda_{\textrm{QCD}}^2x^2))^2}~~,~~~~~~~~~~C_\psi = \frac{96}{\pi^4}
\eeq
where $\Lambda_{\textrm{QCD}}$ is the QCD scale parameter. We note the highly singular nature of the correlator at short distance, especially the $x^8$ dependence. In the evaluation of the Fourier transform and also the sum-rule, this will require a regulator. Moreover, we note that by dimensional grounds the 4-d Fourier transform of $\psi_{\textrm{pert}}$ into momentum space scales as $Q^4$. At first sight, this would seem to imply that the perturbative contribution to the induced Newton constant vanishes, as that contribution is given by the $Q^2$ term in the momentum space function, see Eq.~(\ref{momentumGEuc}). However, that is not correct. When dealing with the physical correlation function, the perturbative result is valid over only part of the $x$ integration region. When treating the perturbative result in only the short distance region, there is a non-zero contribution to the induced Newton constant.

Finally let us discuss the regime of intermediate energies. This can be investigated by means of the Wilson operator product expansion of time-ordered products. For the power suppressed terms in the operator product expansion we borrow from the work that has been performed in sum-rule studies of the correlation function $\psi$ in momentum space. The momentum space correlator of Novikov et al. \cite{Novikov:1979va} is simply related to the Adler-Zee one via
\beq
\Pi_{AZ}(q^2) = \frac{b_0^2}{64 \pi^2}~ \Pi_{NSVZ}(q^2)  \ \ .
\eeq
The work on the operator product expansion of $\Pi_{NSVZ}(q^2)$ in pure QCD ($N_{c} = 3$) is described by~\cite{Bagan:1990sy,Forkel:01}
\bea
\Pi_{NSVZ}(q^2) &=& \left[a_{0} + a_{1}\ln\left(\frac{-q^2}{\mu^2}\right)\right] (-q^2)^{2}\ln\left(\frac{-q^2}{\mu^2}\right)
+ \left[r_{0} + r_{1}\ln\left(\frac{-q^2}{\mu^2}\right)\right] \langle\alpha_{s}F^2\rangle
\nn\\
&+&\, \left[c_{0} + c_{1}\ln\left(\frac{-q^2}{\mu^2}\right)\right]\frac{\langle g F^3\rangle}{(-q^2)}
+ \frac{d_{0}}{(-q^2)^2}\langle\alpha^2_{s}F^4\rangle,
\label{OPE}
\eea
where $\langle (\cdots) \rangle$ are gluon condensate terms
\bea
\langle\alpha_{s}F^2\rangle &=& \langle \alpha_{s}F^{a}_{\mu\nu}F^{a\mu\nu} \rangle
\nn\\
\langle g F^3\rangle &=& \langle g f^{abc}F^{a}_{\mu\nu}F^{b\nu}_{\ \ \rho}F^{c\rho\mu} \rangle
\nn\\
\langle\alpha^2_{s}F^4\rangle &=& 14 \langle (\alpha_{s}f^{abc}F^{a}_{\mu\rho}F^{b}_{\nu}{}^{\rho})^2 \rangle
- \langle (\alpha_{s}f^{abc}F^{a}_{\mu\nu}F^{b}_{\rho\lambda})^2 \rangle
\eea
and the various parameters appearing in Eq.~(\ref{OPE}) are given in terms of the $\alpha_{s} = g^{2}/4\pi$
\bea
a_{0} &=& -2 \left(\frac{\alpha_{s}}{\pi}\right)^{2}\left(1 + \frac{51}{4}\frac{\alpha_{s}}{\pi}\right);
\,\,\,
r_{0} = 4\alpha_{s}\left(1 + \frac{49}{12}\frac{\alpha_{s}}{\pi}\right);
\,\,\,
c_{0} = 8\alpha_{s}^{2};
\,\,\,
d_{0} = 8\pi \alpha_{s}
\nn\\
a_{1} &=& \frac{b_{0}}{2}\left(\frac{\alpha_{s}}{\pi}\right)^{3};
\,\,\,
r_{1} = - b_{0}\frac{\alpha_{s}^{2}}{\pi};
\,\,\,
c_{1} = -58 \alpha_{s}^{3}.
\eea
The first (leading) term is the perturbative contribution. It can be improved by using the renormalization group and asymptotic freedom, which allows a partial resummation of the power series of logarithms appearing in it~\cite{Adler:1982ri}. This will be briefly discussed below. In addition, the position-space forms of the various terms are defined by the Fourier transform of momentum-space results.

\section{Evaluation}

As above we begin our considerations with the long-distance physics. By inserting Eq.~(\ref{infrared}) in the expression for $I_{\textrm{IR}}$ given in the third line of Eq.~(\ref{UV-IR}) and performing the associated integrals, one finds
\beq
I_{\textrm{IR}} = \frac{4 \lambda ^2}{\pi ^2 M_g^4}\,G_{1,3}^{3,0}\left(\frac{M_g^2 t_{0}}{4}\bigg|
\begin{array}{c}
 1 \\
 0,2,3 \\
\end{array}
\right),
\eeq
where
$$
G_{p,q}^{m,n}\left(z\bigg|
\begin{array}{c}
 a_{1},\ldots,a_{n},\ldots,a_{p} \\
 b_{1},\ldots,b_{m},\ldots,b_{q} \\
\end{array}
\right)
$$
is the Meijer G-function~\cite{Prudnikov:86}.

For the UV contribution, we again start our considerations with the perturbative part, which we call $I^{L}_{\textrm{UV}}$. For this we need to regularize the position space integral. We use two methods, which lead to the same result. The QCD scale parameter is given by (at one-loop order)
\bea
&& \Lambda_{\textrm{QCD}}(g(\mu),\mu) = \mu e^{-1/[b g^{2}(\mu^2)]}
\nn\\
&& b = \frac{1}{8\pi^{2}}\left(\frac{11}{3}N_{c} - \frac{2}{3}N_{f}\right)
\eea
where $\mu^{2}$ is an arbitrary subtraction point, $N_{f} = 0$ and $N_{c} = 3$ for gluonic QCD. In this way one gets, with another change of variables
$u = \Lambda_{\textrm{QCD}}^{2} t$
\bea
I^{L}_{\textrm{UV}} &=& C_{\Psi} \Lambda_{\textrm{QCD}}^{2} \int_{0}^{u_0} \frac{du}{u^{2}}\,\frac{\Theta(u)}{(\ln u)^{2}}
\nn\\
\Theta(u) &=& 1 + \sum_{n=1}^{\infty}\sum_{m = 0}^{n} a_{mn}\,\frac{[\ln(\ln u^{-1})]^{m}}{(\ln u^{-1})^{n}}
\eea
The coefficients $a_{mn}$ are loop corrections of higher order. In addition we employ the restriction that $u_{0} = \Lambda_{\textrm{QCD}}^{2} t_{0} < 1$ so that the logarithm $\ln u$ does not vanish in the integration range of $I^{L}_{\textrm{UV}}$.

Let us focus on the leading contribution. We perform the integration by two different ways. First, let us rewrite $I^{L}_{\textrm{UV}}$ as
\beq
I^{L}_{\textrm{UV}} = C_{\Psi} \Lambda_{\textrm{QCD}}^{2} \int^{\infty}_{\ln u_0^{-1}} du\,e^{v}\frac{\Theta(e^{-v})}{(v)^{2}}.
\eeq
In the leading order, $\Theta(e^{-v}) = 1$. Introducing a regulator $e^{-\epsilon v}$ the integral can be easily done to give
\beq
I^{L}_{\textrm{UV}} = C_{\Psi} \Lambda_{\textrm{QCD}}^{2}\,\left\{\frac{e^{x_{0} - x_{0}\epsilon}}{x_{0}} - (\epsilon -1) \Gamma (0,x_{0}(\epsilon -1))\right\},
\eeq
where $x_{0} = \ln u_0^{-1}$ and $\Gamma(a,z)$ is the incomplete gamma function. Taking the limit $\epsilon \to 0^{+}$, one gets
\beq
I^{L}_{\textrm{UV}} = \frac{C_{\Psi} \Lambda_{\textrm{QCD}}^{2}}{x_{0}}\,
\left\{e^{x_{0}} - x_{0} [\text{Chi}(x_{0})+\text{Shi}(x_{0}) + \ln (-x_{0}) - \ln (x_{0})]\right\},
\label{icc}
\eeq
where $\text{Chi}(z)$ ($\text{Shi}(z)$) is the hyperbolic cosine (sine) integral. By choosing the principal branch of the logarithm, one has that $\ln (-x_{0}) - \ln (x_{0}) = i\pi$. Hence taking the real part of $I^{L}_{\textrm{UV}}$ leads us to
\beq
I^{L}_{\textrm{UV}} = \frac{C_{\Psi} \Lambda_{\textrm{QCD}}^{2}}{x_{0}}\,
\left\{e^{x_{0}} - x_{0} [\text{Chi}(x_{0})+\text{Shi}(x_{0})]\right\}.
\label{ic}
\eeq
We observe that $I^{L}_{\textrm{UV}}$ may change sign depending on the values assigned for $x_{0}$.

Now let us calculate $I^{L}_{\textrm{UV}}$ by another method. We follow closely the discussion in~\cite{Adler:1982ri}. First let us consider our calculations in a $2d$-dimensional space, which yields
\beq
I^{L}_{\textrm{UV}} = C_{\Psi} \Lambda_{\textrm{QCD}}^{2} \int^{\infty}_{\ln u_0^{-1}} du\,e^{(d-1)v}\frac{\Theta(e^{-v})}{(v)^{2}}
\eeq
with the contour of integration running along the positive real axis. Let us consider $d$ as a complex parameter; then one is interested in analytically continuing the integral to $d = 2$. When $\textrm{Re}[d] < 1$ and $\textrm{Im}[d] > 0$, the integration contour can be deformed to the contour $C$ depicted in Ref.~\cite{Adler:1982ri}. On the other hand, when $\textrm{Re}[d] < 1$ and $\textrm{Im}[d] < 0$ the contour could be deformed to a contour obtained by reflecting $C$ in the real axis. In this way, the above integral converges for any value of $\textrm{Re}[d]$ and one can analytically continue $\textrm{Re}[d] \to 2$. The regularization prescription must be real, as required by Hermiticity of a quantum field theory; this implies that the limit $d \to 2$ can be prescribed as the average of dimensional continuations to $d = 2 \pm i\epsilon$, with $\epsilon \to 0^{+}$. That means one should take the real part of the evaluation on the contour $C$ alone at the end of the calculations. The inequivalence of the evaluations on both contours is connected to the fact that the analytic continuation of the integral to $\textrm{Re}[d] > 1$ has a branch cut running along the positive real axis from $d = 1$ to infinity.

With the aforementioned prescriptions, let us perform the integral in $I^{L}_{\textrm{UV}}$ at the leading order. We split the contour in two parts $C = C_{1} \cup C_{2}$, with the parametric representations
\bea
z_{1} &=& r e^{i\theta}, 0 < \theta < \frac{\pi}{2}, r = x_{0}
\nn\\
z_{2} &=& r e^{i\theta}, \theta = \frac{\pi}{2}, x_{0} < r < \infty.
\eea
For the first contour $C_{1}$ one obtains
\beq
I^{L}_{\textrm{UV}\,C_{1}} = i \frac{C_{\Psi} \Lambda_{\textrm{QCD}}^{2}}{x_{0}}
\left\{-i x_{0} (d -1) (-\text{Ei}(x_{0} (d -1))+\text{Ei}(i x_{0} (d -1)))+e^{i x_{0} (d -1)}-i e^{x_{0} (d -1)}\right\},
\eeq
where $\text{Ei}(z)$ is the exponential integral function. In turn, for the second contour $C_{2}$ one gets
\beq
I^{L}_{\textrm{UV}\,C_{2}} = -i \frac{C_{\Psi} \Lambda_{\textrm{QCD}}^{2}}{x_{0}}
\left\{e^{i x_{0} (d -1)}+ i x_{0} (d -1) \Gamma (0,-i x_{0} (d -1))\right\}.
\eeq
Hence $I^{L}_{\textrm{UV}} = I^{L}_{\textrm{UV}\,C_{1}} + I^{L}_{\textrm{UV}\,C_{2}}$. Following the above discussed prescription, one arrives at the same result, namely Eq.~(\ref{ic}). Incidentally, as remarked above, within dimensional regularization this leading contribution vanishes when one performs the integral for all values of $t$. Hence our result should be zero for $t_0 \to \infty$, or $x_{0} \to -\infty$. This is precisely what happens when one takes the limit $x_{0} \to -\infty$ of Eq.~(\ref{icc}).

Now let us calculate the contribution of the non-perturbative terms coming from the operator product expansion of $\Psi(x^2)$, which are the second and the third terms of Eq.~(\ref{OPE}). Using the Fourier transform to identify the position space correlation function, one finds
\beq
\Psi^{OPE}_{UV}(x^2) = \frac{b^{2}_{0}}{256\pi^{4}}\left\{- \frac{4 r_{1}}{(x^2)^2}\langle\alpha_{s}F^2\rangle +
\frac{\langle g F^3\rangle}{x^2}\left[c_{0} - c_{1}\left(\ln\left(\frac{x^{2}\mu^{2}}{4}\right) + 2 \gamma\right)\right]\right\},
\label{OPE2}
\eeq
where $\gamma =  0.5772$ is the Euler-Mascheroni constant. Hence
\beq
I^{OPE}_{UV} = \frac{b^{2}_{0} \alpha_{s}^{2}}{256\pi^{4}}\left\{\frac{4 b_{0} t_{0}}{\pi}  \langle\alpha_{s}F^2\rangle
+ \frac{t^{2}_{0}}{2} \left[4(2  + 29 \gamma \alpha_{s})
+ 29\alpha_{s}\left(\ln\left(\frac{\mu^4 t_{0}^2}{16}\right)-1\right)\right] \langle g F^3\rangle \right\}.
\eeq
Finally, collecting our results and recalling that $x_{0} = \ln u_0^{-1} = - \ln(\Lambda_{\textrm{QCD}}^{2} t_{0})$, one has that
\bea
\hspace{-1cm}
\frac{1}{16\pi G_{\textrm{ind}}} &=& - \frac{\pi^2}{96} \left\{\frac{4 \lambda ^2}{\pi ^2 M_g^4}\,G_{1,3}^{3,0}\left(\frac{{\bar M}_{g}^2\,e^{-x_{0}}}{4}\bigg|
\begin{array}{c}
 1 \\
 0,2,3 \\
\end{array}
\right)
+ \frac{C_{\Psi} \Lambda_{\textrm{QCD}}^{2}}{x_{0}}\,
\left[e^{x_{0}} - x_{0} \Bigl(\text{Chi}(x_{0})+\text{Shi}(x_{0})\Bigr)\right] \right.
\nn\\
&+&\,\left. \frac{b^{2}_{0} \alpha_{s}^{2}}{256\pi^{4}}\left[\frac{4 b_{0} e^{-x_{0}}}{\Lambda_{\textrm{QCD}}^{2}\pi} \langle\alpha_{s}F^2\rangle
+ \frac{e^{-2 x_{0}}}{2\Lambda_{\textrm{QCD}}^{4}} \left(4(2  + 29 \gamma \alpha_{s})
+ 29\alpha_{s}\left(\ln\left(\frac{\mu^4 e^{-2 x_{0}}}{16\Lambda_{\textrm{QCD}}^{4}}\right)-1\right)\right) \langle g F^3\rangle \right]
\right\},
\label{g-final}
\eea
where ${\bar M}_{g} = M_g/\Lambda_{\textrm{QCD}}$.

\begin{figure}[htb]
\begin{center}
\includegraphics[height=120mm,width=130mm]{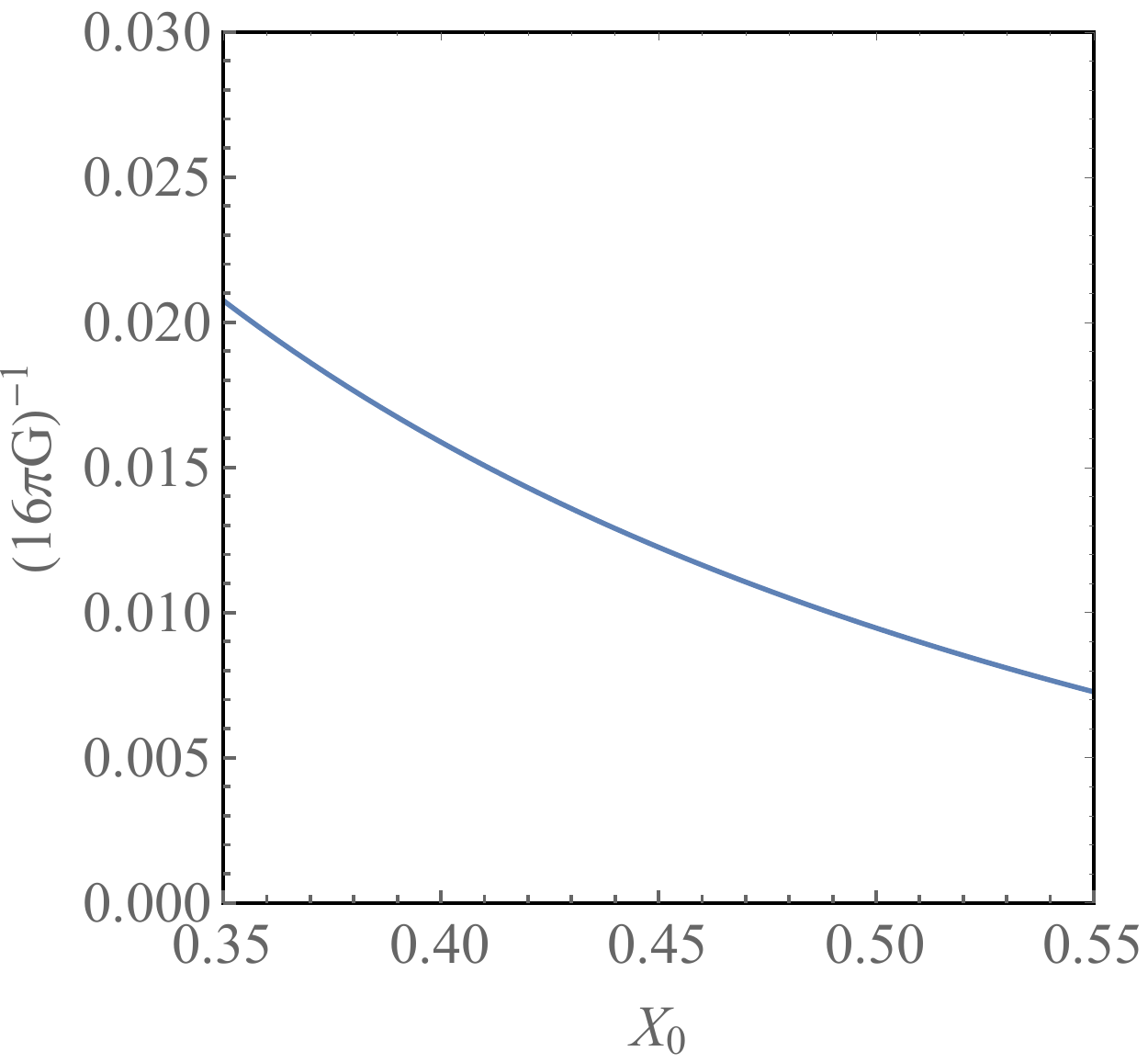}
\caption{$G_{\textrm{ind}}^{-1}$ as a function of {\color{red} the matching scale }$X_{0}$. The values of $X_0$ are in units of GeV$^{-1}$, and the vertical axis is in units of GeV$^2$.}
\label{grec}
\end{center}
\end{figure}

Let us now perform a numerical analysis of the result~(\ref{g-final}). In evaluating the sum rule, we use the lattice data given by Ref.~\cite{Chen:2005mg}. The scalar glueball mass found there is
\beq
M_g =1.71 \pm 0.05 \pm0.08 ~~{\rm GeV} \  \ ,
\eeq
and the glueball coupling is
\beq \label{gluecoupling}
\lambda = 1.1 \pm 0.22  ~~{\rm GeV}^3  \ \ .
\eeq
The mass is consistent with many previous investigations. The glueball coupling turns out to be almost four times larger than found in a previous related study \cite{Liang:1992cz}. Because the glueball contribution to the induced Newton constant is negative, using a smaller coupling would make the final answer more positive. However, we see no reason not to use the most recent value as the study of \cite{Chen:2005mg} is a significant advance over previous work. On the other hand, for the OPE coefficients we employ the following given values of the parameters~\cite{Forkel:01,Chen:2005mg}:
\bea
\langle\alpha_{s}F^2\rangle &=& 0.04 \, \textrm{GeV}^{4}
\nn\\
\langle g F^3\rangle &=& -1.5 \langle\alpha_{s}F^2\rangle^{3/2}
\nn\\
\mu &=& 2 \, \textrm{GeV}
\nn\\
\alpha_{s} &=& 0.2.
\eea
Fig.~\ref{grec} illustrates the induced Newton's constant as a function of $X_{0} = \sqrt{t_{0}}$. The distance interval shown in the plot corresponds to the range $X_0^{-1} = 1.8$~GeV on the right hand side to $X_0^{-1} = 2.8$~GeV on the left. Because the lattice calculation utilises a scale of 2 GeV, and reveals a glueball of 1.7 GeV but does not investigate higher states, we quote our result for a matching scale of $X_0^{-1} = 2$~GeV:
\beq
\frac{1}{16\pi G_{\textrm{ind}}} = 0.0095 \pm 0.0030~ {\rm GeV}^2
\eeq
Our error bar is determined by examining changes in the input parameters, with the most sensitive being the glueball coupling $\lambda$ of Eq.~(\ref{gluecoupling}).

{\color{red} As Fig.~\ref{grec} clearly reveals, $G$ is not absolutely constant in the energy range considered, which suggests that there exists some residual scale dependence in our evaluation}. The matching at the scale $X_0$ is not perfect at the energies which we are working. This could be explained by the fact that the lattice data only reveals one glueball mass and coupling. When matching at $X_0^{-1} \sim 2$~GeV, this should capture the bulk of the long distance effect as the glueball mass is 1.7 GeV. However, when the matching takes place at higher energy, the presence of extra glueball excited states probably would be relevant. Because the glueball contribution is negative, this would have the effect of decreasing the result at short distances, going in the direction of making the matching more independent of the scale. We do not have a numerical evaluation of this physics, but at least the direction of the effect is correct.

\section{Discussion}

The evaluation of the Adler-Zee sum-rule is an exercise in the study of QCD, but one which may have some implications for gravitational theories. The resultant shift in the effective Planck mass is by an amount which is typical of the QCD scale (of course) and positive. The expected scaling of the various ingredients are such that the result would stay positive for SU(N) theories with larger values of N.

An early previous evaluation \cite{Krasnikov:1985ar} also yielded a positive value of the induced Planck mass. This evaluation subtracted off the perturbative result, defined a finite energy sum-rule for the remainder, and modeled the spectrum. Our result uses lattice data rather than models and more properly includes the perturbative contribution.

In addition to the induced contribution to the Planck mass, we expect a contribution to the cosmological constant, given in Eq.~(\ref{cosmconst}). The standard value for the gluon condensate yields a value of
\beq
\Lambda_{\textrm{ind}} = -0.0034 ~{\rm GeV}^4    \ \ .
\eeq
for two light quark flavors. {\color{red} However, the value of the gluon condensage is not firmly known. Indeed, Holdom has given arguments that the gluon condensate could vanish, drawing attention to the absence of both experimental and theoretical evidence for a nonvanishing gluon condensate in massless QCD~\cite{Holdom:2007gg}. Some of the difficulty in direct lattice calculations is the presence of a dimensionful cut-off for the lattice, and also disentangling the gluonic contribution from that of massive quarks. If this assertion is correct, it would have important implications for the use of induced effects in gravity theories}. Our induced shift in the Planck mass remains positive and within the quoted error bars if we set the gluon condensate to zero.

It is evident from the derivation that shift in the value of $G$ due to QCD is only valid for wavelengths greater than the QCD scale. For shorter wavelengths the effect is different, and the effective value of $G$ above the QCD scale will be different. Moreover the effect on the graviton propagator, defined by the graviton two-point function, will be a more complicated function of the momentum, including the development of imaginary component, bringing in the Lee-Wick mechanism  \cite{Lee:1969fy, Tomboulis:1977jk, Grinstein:2008bg}. While the case of QCD is not likely to be of phenomenological importance, because the QCD scale is so far below the Planck scale, if there are other strongly interacting gauge theories at much higher energy, there could be important consequences.
Some recent suggestions include the interactions of the spin connection \cite{Donoghue:2016vck} and even of the gravitational field itself \cite{Holdom:2015kbf}.
These effects deserve further study.

\section*{Acknowledgements} We thank Bob Holdom, Jing Ren, Pierre Vanhove, Keh-Fei Liu and Claus Kiefer for useful discussions. This work has been supported in part by the National Science Foundation under grant NSF PHY15-20292.

\section*{Appendix - The cosmological constant sum rule}

In this Appendix we elucidate the mismatch involving a two-point representation for the cosmological constant, where the works of Novikov et al (NSVZ) \cite{Novikov:1979va} and of Brown and Zee \cite{Brown:1982am} yield sum-rules that differ by a factor of two. For QCD, the  correct one is that of
NSVZ, which reads
\beq\label{correct}
\Lambda_{\textrm{ind}} =
- \frac{i}{16} \int d^{4} z \langle T \{{\bar T}(z){\bar T}(0)\}\rangle = -\frac{b_0}{32}\left\langle \frac{\alpha_{s}}{\pi}\,F^{a}_{\mu\nu}F^{a\mu\nu} \right\rangle \ \ .
\eeq
where
$$
b_0 = \frac{11}{3}N_{c} - \frac{2}{3}N_{f}.
$$
Here $N_{f}$ is the number of quark species in the theory. For purely gluonic QCD one has $N_{f} = 0$ and $N_{c} = 3$.

The equality on the right side is the sum-rule identity given by NSVZ, and we have used Eq.~(\ref{cosmconst}) to relate that result to $\Lambda_{\textrm{ind}}$ in order to obtain the equality on the left side. The sum-rule of Brown and Zee corresponds to the left-hand equality but with a coefficient that is twice as large, with the $i/16$ being replaced by $i/8$.

The issue hinges on the two-graviton coupling called  $\tau_{\mu\nu,\alpha\beta}(x) $ in Eq.~(\ref{expansion}). When Brown and Zee expand the action they include only the linear coupling $h^{\mu\nu}T_{\mu\nu}$. By matching the resultant Lagrangian to that of a cosmological constant, their sum-rule is obtained. We can see how this is changed by including the two graviton coupling.
 First let us exhibit the representations for $\Lambda_{\textrm{\textrm{ind}}}$. Recalling that $h_{\mu\nu} = (1/4)\eta_{\mu\nu}h$, one has that $\sqrt{-g} = 1 + (1/2)h + (1/16)h^2$. Since $\sqrt{-g}{\cal L}^{(0)}_{\textrm{eff}}[g_{\mu\nu}]
= -\sqrt{-g}\Lambda_{\textrm{ind}}$, comparison with Eq.~(\ref{Leff}) leads to the following representations for the cosmological constant:
\bea
\Lambda_{\textrm{ind}} &=& \frac{1}{4}\langle T(x) \rangle
\nn\\
\Lambda_{\textrm{ind}} &=&- \frac{1}{4} \eta^{\mu\nu}\eta^{\rho\sigma}\langle \tau_{\mu\nu,\rho\sigma}(x) \rangle
- \frac{i}{8} \int d^{4} z \langle T \{{\bar T}(z){\bar T}(0)\}\rangle.
\label{induced2}
\eea
The second of these is the correct sum-rule in a generic theory. The result of Brown and Zee is obtained if one drops $\tau_{\mu\nu,\rho\sigma}$. However, in Yang-Mills theories, the two-graviton coupling does contribute, and resolves the discrepancy in the sum-rules. The Yang-Mills Lagrangian in a generic curved background is given by
\beq
\sqrt{-g}\,{\cal L}_{YM} = \sqrt{-g}\left[-\frac{1}{4} F^{a}_{\mu\nu}F^{a\mu\nu}\right],
\eeq
where the index $a$ is summed over the generators of the gauge group $G$. The field strength is given by
\beq
F^{a}_{\mu\nu} = \partial_{\mu} A^{a}_{\nu} - \partial_{\nu} A^{a}_{\mu} + g f^{abc} A^{b}_{\mu} A^{c}_{\nu}
\eeq
where $f^{abc}$ are the structure constants of $G$. For weak fields the metric can be expanded using $g_{\mu\nu} = \eta_{\mu\nu} + h_{\mu\nu}$. One finds that
\beq
\sqrt{-g}\,{\cal L}_{YM} = -\frac{1}{4} F^{a}_{\mu\nu}F^{a\mu\nu}
- \frac12 h^{\mu\nu} T_{\mu\nu} + \frac14 h^{\mu\nu}h^{\alpha\beta}\tau_{\mu\nu,\alpha\beta}
+ {\cal O}(h^3)
\eeq
where the energy-momentum tensor for the Yang-Mills field in Minkowski space-time is given by
\beq
T_{\mu\nu} = - F^{a}_{\lambda\mu}F^{a\lambda}_{\ \ \nu}
+ \frac{1}{4}\eta_{\mu\nu} F^{a}_{\alpha\beta}F^{a\alpha\beta}.
\eeq
In addition, the tensor $\tau_{\mu\nu,\alpha\beta}$ reads
\beq
\tau_{\mu\nu,\alpha\beta} =
- 2\eta_{\alpha\nu}F_{a\lambda\mu}F^{a\lambda}_{\ \ \ \beta} - F^{a}_{\mu\alpha}F^{a}_{\nu\beta}
+ \frac{1}{4}{\cal P}_{\alpha\beta\mu\nu} F^{a}_{\gamma\delta}F^{a\gamma\delta}
+ \eta_{\alpha\beta}F^{a}_{\kappa\mu}F^{a\kappa}_{\ \ \ \nu}
\eeq
where
\beq
{\cal P}_{\alpha\beta\mu\nu} = \frac{1}{2}(\eta_{\alpha\mu}\eta_{\beta\nu} + \eta_{\alpha\nu}\eta_{\beta\mu}
-\eta_{\alpha\beta}\eta_{\mu\nu}).
\eeq
Now let us prove the result~(\ref{correct}). In order to calculate $\Lambda_{\textrm{ind}}$, one needs an expression for $T(x)$. With the introduction of a dynamical scale-invariance breaking, one gets $T(x) \neq 0$. This is the well-known trace anomaly. The trace anomaly formula for pure QCD is given by [see Ref.~\cite{Adler:1982ri} and references cited therein]
\beq
T(x) = \frac{\beta(g)}{2g} F^{a}_{\mu\nu}F^{a\mu\nu},
\label{trace-qcd}
\eeq
where $\beta(g)$ is the renormalization-group beta function, which in the lowest order is given by
\beq
\beta(g) = -\frac{1}{2}\,b g^{3}.
\eeq
In this way one also finds
\beq
\eta^{\mu\nu}\eta^{\rho\sigma}\tau_{\mu\nu\rho\sigma} =
\frac{\beta(g)}{2g} F^{a}_{\gamma\delta}F^{a\gamma\delta} = T(x).
\label{trace-qcd2}
\eeq
In Eqs.~(\ref{trace-qcd}) and~(\ref{trace-qcd2}) it is to be understood that $F^{a}_{\mu\nu}$ is the renormalized field strength. From the first of the relations presented in Eq.~(\ref{induced2}), one gets
\beq
\Lambda_{\textrm{ind}} = -\frac{b_0}{32}\left\langle \frac{\alpha_{s}}{\pi}\,F^{a}_{\mu\nu}F^{a\mu\nu} \right\rangle.
\eeq
Now let us discuss the two-point representation for the induced cosmological constant. Using the sum rule derived in Ref.~\cite{Novikov:1979va}, which states that
\beq
i \int dz \langle T \{{\bar T}(z){\bar T}(0)\}\rangle =
\frac{b_0}{2}\left\langle \frac{\alpha_{s}}{\pi}\,F^{a}_{\mu\nu}F^{a\mu\nu} \right\rangle
\eeq
it is easy to see that the second of the relations presented in Eq.~(\ref{induced2}) produces the same result for $\Lambda_{\textrm{ind}}$. This proves our assertion.

We do not evaluate the cosmological constant sum-rule numerically because of the possibility of delta function OPE contributions to the position space sum-rule \cite{Novikov:1983gd}. Because of the extra powers of $x^2$, these do not influence the Adler-Zee formula, but they would enter into the cosmological constant sum-rule. For the cosmological constant, Eq.~(\ref{cosmconst}) is still the most reliable estimate.

\end{document}